\documentclass[
 reprint,
superscriptaddress,
 amsmath,amssymb,
 aps,
pra,
]{revtex4-1}
\usepackage{hyperref}
\newcommand{\bi}{\mathbf}
\usepackage{graphicx}
\usepackage{dcolumn}
\usepackage{bm}
\usepackage{verbatim}

\usepackage{color}

\begin{document}


\title{Quantum Electrodynamical Bloch Theory with Homogeneous Magnetic Fields}

\author{Vasil~Rokaj}
\email{vasil.rokaj@mpsd.mpg.de}
\affiliation{Max Planck Institute for the Structure and Dynamics of Matter,
Center for Free Electron Laser Science, 22761 Hamburg, Germany}

\author{Markus~Penz}
\affiliation{Max Planck Institute for the Structure and Dynamics of Matter,
Center for Free Electron Laser Science, 22761 Hamburg, Germany}

\author{Michael~A.~Sentef}
\affiliation{Max Planck Institute for the Structure and Dynamics of Matter,
Center for Free Electron Laser Science, 22761 Hamburg, Germany}
\author{Michael~Ruggenthaler}

\affiliation{Max Planck Institute for the Structure and Dynamics of Matter,
Center for Free Electron Laser Science, 22761 Hamburg, Germany}

\author{Angel~Rubio}
\email{angel.rubio@mpsd.mpg.de}
\affiliation{Max Planck Institute for the Structure and Dynamics of Matter,
Center for Free Electron Laser Science, 22761 Hamburg, Germany}
\affiliation{Center for Computational Quantum Physics (CCQ), Flatiron Institute, 162 Fifth Avenue, New York NY 10010}


\date{\today}

\begin{abstract}
 We propose a solution to the problem of Bloch electrons in a homogeneous magnetic field by including the quantum fluctuations of the photon field. A generalized quantum electrodynamical (QED) Bloch theory from first principles is presented. In the limit of vanishing quantum fluctuations we recover the standard results of solid-state physics, the fractal spectrum of the Hofstadter butterfly. As a further application we show how the well known Landau physics is modified by the photon field and that \textit{Landau polaritons} emerge. This shows that our QED-Bloch theory does not only allow to capture the physics of solid-state systems in homogeneous magnetic fields, but also novel features that appear at the interface of condensed matter physics and quantum optics.   
\end{abstract}

\pacs{Valid PACS appear here}
\maketitle


Cavity QED materials is a growing research field bridging quantum optics~\cite{Cohen, Gerry}, polaritonic chemistry~\cite{Ebbesen, Ribeiro, RuggiNature, Feist, Johannes PNAS 2017}, and materials science, such as light-induced new states of matter achieved with classical laser fields~\cite{Basov, Buzzi}. Photon-matter interactions have recently been suggested to modify electronic properties of solids, like superconductivity and electron-phonon coupling~\cite{Sentef, Schlawin, Curtis, Imamoglou, Kavokin}. On the other hand, materials in classical magnetic fields are known to give rise to several novel phenomena like the Landau levels~\cite{Landau}, the integer~\cite{Klitzing, Kohmoto} and the fractional quantum Hall effect~\cite{Laughlin}, and the Hofstadter butterfly~\cite{Hofstadter} which can be now accessed experimentally with high resolution~\cite{Dean, Wang, Forsythe}. An open question in this field is whether Bloch theory is applicable for solids in the presence of a homogeneous magnetic field, which breaks translational symmetry. This issue was solved partially by introducing the magnetic translation group, which however puts fundamental limitations on the allowed strength of the magnetic field, since it permits only rational fluxes through the unit cell~\cite{Kohmoto, Zak, J.Zak}. 

\begin{figure}[h]
\includegraphics[width=\columnwidth]{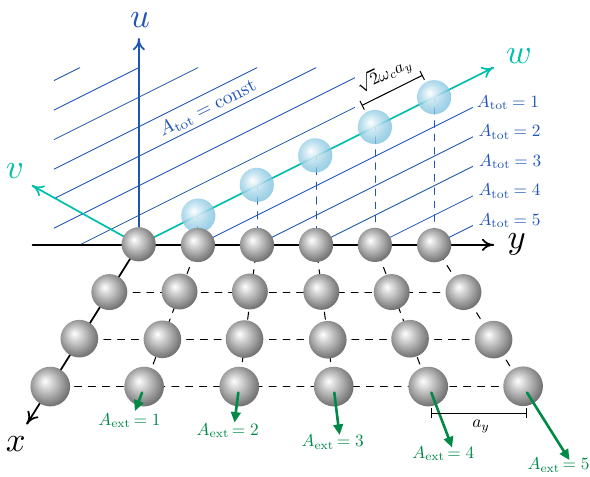}
\caption{\label{figure}Here $\mathbf{A}_{\textrm{ext}}$ breaks periodicity along $y$ of an otherwise periodic material in the $(x,y)$ plane with lattice constant $a_{y}$. Including the quantized field $\hat{\mathbf{A}}$ proportional to the photonic coordinate $u$, we obtain the total vector potential $\hat{\mathbf{A}}_{\textrm{tot}}=\hat{\mathbf{A}}+\mathbf{A}_{\textrm{ext}}$ which is constant in the polaritonic direction $w$ which makes the system periodic along $w$ with lattice constant $\sqrt{2} \omega_c a_{y}$, where $\omega_c=eB/m_{\textrm{e}}$. Thus, when embedding the $(x,y)$ plane into the higher-dimensional space involving the coordinate $u$, periodicity gets restored, while in the electronic subspace the system is aperiodic.}
\end{figure}

\begin{figure}[h]
\includegraphics[width=\columnwidth]{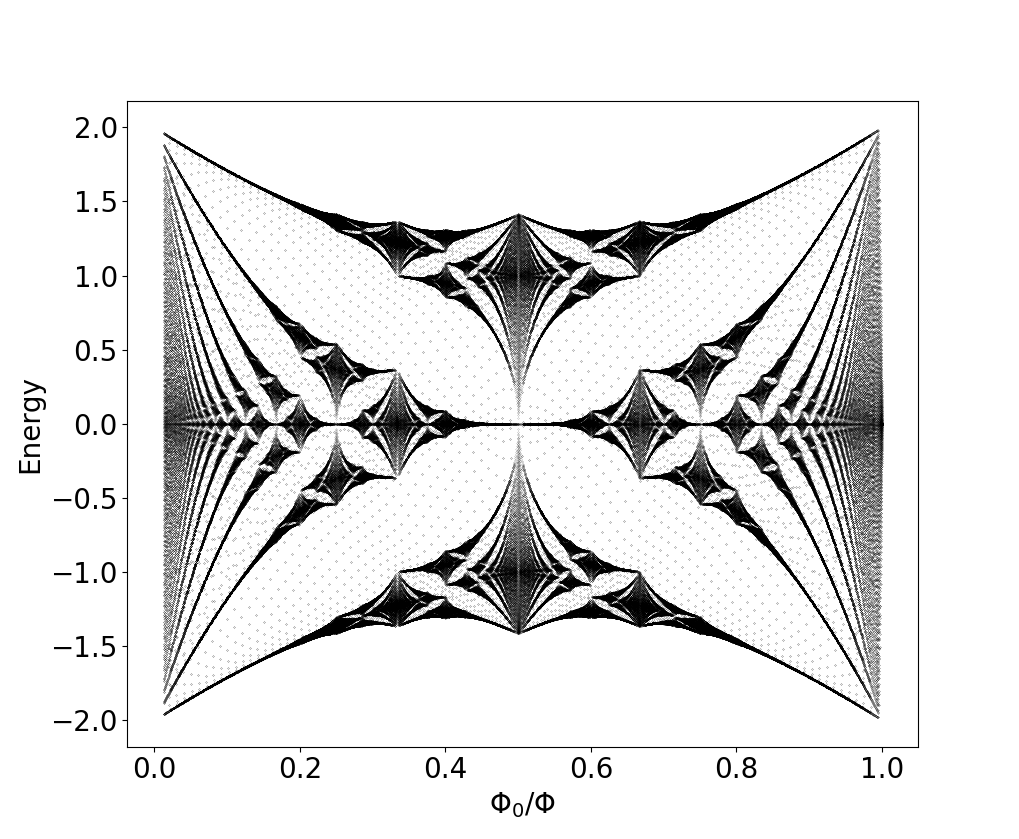}
\caption{Energy spectrum of a 2D solid in a perpendicular homogeneous magnetic field as a function of the inverse relative flux $\Phi_0/\Phi=\hbar /\textrm{e}Ba_xa_y$.}
\label{Butterfly}
\end{figure}

\begin{figure}[h]
\includegraphics[width=\columnwidth]{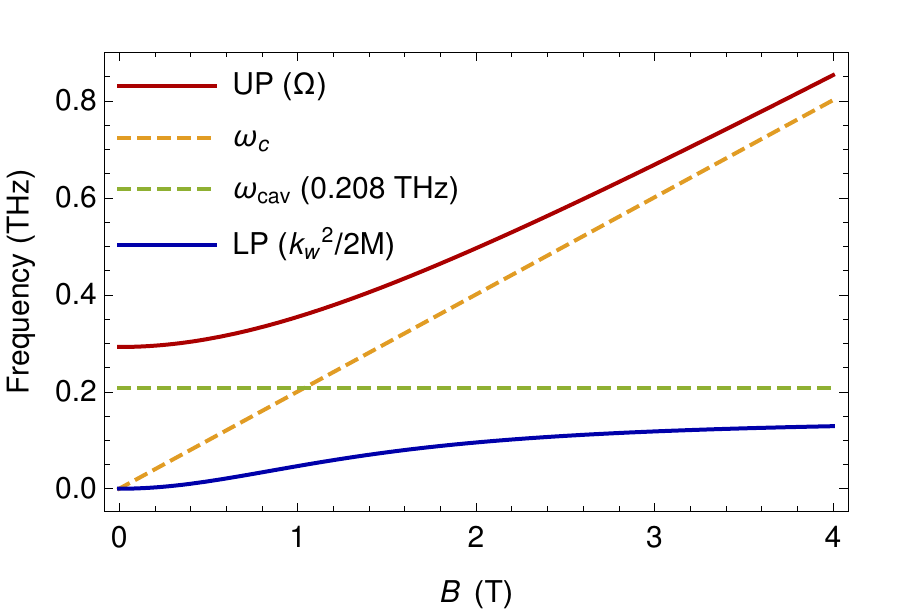}
\caption{Upper (red line) and lower (blue line) polaritonic excitations of~(\ref{LandauFl}) as a function of the strength of the magnetic field $B$ (T). The upper polariton (UP) asymptotically reaches the dispersion of the cyclotron transition $\omega_c=eB/m^*$ (orange dashed line). The lower polariton (LP) does not reach the empty cavity frequency $\omega_{\textrm{cav}}$~\cite{Keller}.}
\label{Landau}
\end{figure}

In this Letter, by combining QED with solid-state physics, we provide a consistent and comprehensive theory for solids interacting with homogeneous electromagnetic fields, both classical and quantum, in which a magnetic field of arbitrary strength can be treated non-perturbatively. Our main findings are: (i) The quantum fluctuations of the electromagnetic field allow to restore translational symmetry that is broken due to an external homogeneous magnetic field (see Fig.~\ref{figure}). 
 (ii) We generalize Bloch theory and provide a Bloch central equation for solids in the presence of a homogeneous magnetic field and its quantum fluctuations. (iii) Applying our framework for a 2D solid in a perpendicular homogeneous magnetic field, in the limit of no quantum fluctuations, we recover the Hofstadter butterfly (see Fig.~\ref{Butterfly}). (iv) For a 2D electron gas in a cavity and under the influence of a perpendicular homogeneous magnetic field we find \textit{Landau polaritons}~\cite{Kono, Faist, Keller}. The spectrum of the Landau polaritons (in atomic units) is
\begin{eqnarray}\label{LandauFl}
E_{j,k_w}=k^2_w/2M+\Omega\left(j+1/2\right).
\end{eqnarray}
The frequency of the upper polariton is $\Omega=\sqrt{\omega^2_c+\omega^2_p}$ and depends on the cyclotron frequency $\omega_c=eB/m_{\textrm{e}}$ and the local electron density $n_{\textrm{e}}$ via $\omega_p=\sqrt{n_{\mathrm{e}} e^2/m_{\textrm{e}}\epsilon_0}$. The kinetic energy $k^2_w/2M$ corresponds to the lower polariton (see Fig.~\ref{Landau}) and will be explained in what follows.

\textit{Translational Symmetry with Homogeneous Magnetic Fields}.---Non-relativistic QED describes electrons minimally coupled to the electromagnetic field, both classical and quantum. For the description of the photon field we follow the standard procedure of assuming a finite box of volume $V$~\cite{Cohen, spohn2004, Gerry}. In the usual case of a solid, the volume $V$ does not constitute a physical quantity. In this case the local electron density $n_{\textrm{e}}=N/V$ is the quantity to work with, since the volume $V$ and the number of electrons $N$ tend to infinity in such a way that $n_{\textrm{e}}$ is constant. On the other hand, if we consider a solid confined in a cavity, the mode volume determines the coupling of the cavity modes to the electrons~\cite{Gerry, Ebbesen, Feist, RuggiNature, Johannes PNAS 2017} and the volume becomes a physical quantity. Our starting point in both cases is the Pauli-Fierz Hamiltonian in the single mode limit~\cite{Cohen, spohn2004, Gerry, Rokaj} 
\begin{eqnarray}\label{velgauge} 
\hat{H}&=&\sum\limits^{N}_{j=1}\left[\frac{1}{2m_{\textrm{e}}}\left(\mathrm{i}\hbar \mathbf{\nabla}_{j}+e\hat{\mathbf{A}}(\bi{r}_j)+e\mathbf{A}_{\textrm{ext}}(\bi{r}_j)\right)^2 +v_{\textrm{ext}}(\mathbf{r}_{j})\right]\nonumber\\
&+&\frac{1}{4\pi\epsilon_0}\sum\limits^{N}_{j< k}\frac{e^2}{|\mathbf{r}_j-\mathbf{r}_k|}+\hbar\omega\left(\hat{a}^{\dagger}\hat{a}+\frac{1}{2}\right),
\end{eqnarray}
Here $\mathbf{A}_{\textrm{ext}}(\mathbf{r})$ is an external vector potential. Being interested in the case of a homogeneous magnetic field we choose $\mathbf{A}_{\textrm{ext}}(\mathbf{r})$ in Landau gauge, $\mathbf{A}_{\textrm{ext}}(\mathbf{r})=-\mathbf{e}_x B y$~\cite{Landau}, which gives rise to a constant magnetic field in the $z$-direction, $\mathbf{B}_{\textrm{ext}}=\nabla \times \mathbf{A}_{\textrm{ext}}(\mathbf{r})=\mathbf{e}_z B$.

Moreover, $\hat{\mathbf{A}}(\bi{r})$ is the quantized vector potential of the electromagnetic field in Coulomb gauge~\cite{spohn2004}
\begin{equation}\label{eq2.4b}
\hat{\bi{A}}(\bi{r})=\left(\frac{\hbar}{\epsilon_0 V}\right)^{\frac{1}{2}}\frac{\bm{\epsilon}}{\sqrt{2\omega}}\left[ \hat{a}e^{\mathrm{i}\bm{\kappa}\cdot\bi{r}}+\hat{a}^{\dagger}e^{-\mathrm{i}\bm{\kappa}\cdot\bi{r}}\right].
\end{equation}
Here $\bm{\kappa}$ is the wave vector, $\omega=c|\bm{\kappa}|$ is the frequency, and $\bm{\epsilon}$ is the transversal polarization vector~\cite{Cohen, Gerry, spohn2004}. The annihilation and creation operators in terms of the displacement coordinates $q$ and their conjugate momenta $\partial_q=\partial/\partial q$ are  $\hat{a}=\left(q+\partial_q\right)/\sqrt{2}$ and $\hat{a}^{\dagger}=\left(q-\partial_q\right)/\sqrt{2}$. The quantized field in our theory captures the back-reaction of matter to the electromagnetic field. For that purpose we choose the quantized field and the external field to have same polarization, $\bm{\epsilon}=\mathbf{e}_x$. Such back-reactions are essential in solid-state physics, e.g., in the semi-classical microscopic-macroscopic connection that determines the induced fields inside a material~\cite{Mochan, Maki, Ehrenreich}. In cavity QED these back-reactions get enhanced by cavity confinement, and in this case the quantized field models the influence of the cavity modes. 

In Bloch theory~\cite{Mermin} the external potential is assumed periodic, $v_{\textrm{ext}}(\mathbf{r})=v_{\textrm{ext}}(\mathbf{r}+\mathbf{R}_{\mathbf{n}})$, where $\mathbf{R}_{\mathbf{n}}$ is a Bravais lattice vector. To analyze conveniently the external vector potential we choose the lattice vectors $\mathbf{R}_{\mathbf{n}}=na_x\mathbf{e}_x+ma_y\mathbf{e}_y+la_z\mathbf{e}_z$. Having a periodic external potential and a uniform magnetic field, one would expect a periodic solution using Bloch theory. Yet, it is obvious that $\mathbf{A}_{\textrm{ext}}(\mathbf{r})$ breaks translational symmetry since it is linear in $y$. The quantized vector potential~(\ref{eq2.4b}) is not invariant under the translation $\mathbf{r}\rightarrow \mathbf{r}+\mathbf{R}_{\mathbf{n}}$ either. As a consequence the Pauli-Fierz Hamiltonian~(\ref{velgauge}) is not periodic and Bloch's theorem is not applicable. 

We propose that the problem of broken translational symmetry can be resolved in the \textit{optical limit}. Therein the quantized vector potential is assumed uniform and has no spatial dependence and as a consequence $\hat{\bi{A}}=\mathbf{e}_x q \sqrt{\hbar /\epsilon_0 V\omega}$. But what exactly does the optical limit mean for a solid?  The optical limit is valid when the wavelength of the electromagnetic field is much larger than the size of the electronic system. But solids compared to the size of an atom are infinitely large systems, especially in Bloch theory where full periodicity is assumed. This implies that in the optical limit the wavelength of the field should be infinite and the frequency should tend to zero. Naively, taking $\omega\rightarrow 0$ in $\hat{\bi{A}}$ seems to lead to divergencies in~(\ref{eq2.4b}). However, if the limit is performed consistently by taking into account the back-reaction of matter due to the square of the vector potential, no divergencies arise. 

To that end, we isolate the purely photonic part of $\hat{H}$ which includes the bare photon mode $\omega$ plus the square of the vector potential $\hat{H}_p=\hbar\omega\left(\hat{a}^{\dagger}\hat{a}+1/2\right)+\hat{\mathbf{A}}^2Ne^2/2m_{\textrm{e}}$. In terms of the photonic coordinate $q$ and its momentum $\partial_q$ it is $\hat{H}_p=\hbar\omega/2\left(-\partial^2_q+q^2\right)+q^2Ne^2\hbar/2m_{\textrm{e}}\omega\epsilon_0 V$. Introducing the dressed frequency $\tilde{\omega}^2=\omega^2+\omega^2_p$ and the coordinate $u=q\sqrt{\tilde{\omega}/\omega}$, $\hat{H}_p$ takes the form $\hat{H}_p=\hbar\tilde{\omega}/2\left(-\partial^2_u+u^2\right)$ where the frequency $\omega_p$ depends on the electron density $n_{\mathrm{e}}$ and is given by $\omega_p=\sqrt{n_{\mathrm{e}} e^2/m_{\textrm{e}}\epsilon_0}$. The frequency $\omega_p$ is a diamagnetic shift induced by the collective coupling of the electrons to the transversal photon field~\cite{Rokaj, Todorov, TodorovSirtori}. The vector potential as a function of $u$ is $\hat{\mathbf{A}}=u\mathbf{e}_x\sqrt{\hbar/\epsilon_0 V\tilde{\omega}}$. In the optical limit the dressed frequency $\tilde{\omega}$ goes to $\omega_p$ and substituting $\hat{H}_p$ and $\hat{\mathbf{A}}$ back into~(\ref{velgauge}) we obtain the Hamiltonian in the optical limit
\begin{eqnarray}\label{optical}
&&\hat{H}_{\textrm{opt}}=\sum^{N}_{j=1}\left[-\frac{\hbar^2}{2m_{\textrm{e}}}\nabla^2_j+\frac{\textrm{i}\hbar e}{m_{\textrm{e}}}\left(\hat{\mathbf{A}}+\mathbf{A}_{\textrm{ext}}(\mathbf{r}_j)\right)\cdot\nabla_j +v_{\textrm{ext}}(\mathbf{r}_j)\right]\nonumber\\
&&+\frac{1}{4\pi\epsilon_0}\sum\limits^{N}_{j< k}\frac{e^2}{|\mathbf{r}_j-\mathbf{r}_k|}+\frac{e^2}{2m_{\textrm{e}}}\sum^{N}_{j=1}\left(\hat{\mathbf{A}}+\mathbf{A}_{\textrm{ext}}(\textbf{r}_j)\right)^2-\frac{\hbar\omega_p}{2}\partial^2_u\nonumber\\
\end{eqnarray}
The quantized vector potential in the optical limit is $\hat{\mathbf{A}}=\mathbf{e}_xu\sqrt{\hbar /\epsilon_0V\omega_p}$. For a periodic potential $\hat{H}_{\textrm{opt}}$ is still not periodic in the electronic coordinates because $\mathbf{A}_{\textrm{ext}}(\mathbf{r})$ is linear in $y$. But the optical Hamiltonian $\hat{H}_{\textrm{opt}}$ is periodic under the generalized translation
\begin{eqnarray}\label{symmetry}
(\mathbf{r}_j,u)\longrightarrow\left(\mathbf{r}_j+\mathbf{R}_{\mathbf{n}},u+Bma_y\sqrt{\epsilon_0 V\omega_p/\hbar}\right).
\end{eqnarray}
This proves our claim that in the optical limit the broken translational symmetry, caused by the homogeneous magnetic field, gets restored (see Fig.~\ref{figure}). 

\textit{QED-Bloch Theory with Homogeneous Magnetic Fields}.---Having restored translational symmetry we can derive a Bloch central equation for solids in homogeneous magnetic fields. Instead of expressing the unfeasible many-electron interacting problem of Eq.~(\ref{optical}), we will employ the independent electron approximation which resembles the usual approach of density-functional theory (DFT). Such an approach is consistent with Bloch theory, which is not a theory of one electron in a periodic potential, but of many non-interacting electrons. Thus, to account for the collective coupling of the electrons to the photon field, we use an effective electron density to capture the back-reaction correctly. Any further exchange and correlation effects would need the inclusion of effective fields as introduced in quantum-electrodynamical DFT~\cite{RuggiNature, RuggiArxiv}. Introducing the cyclotron frequency the Hamiltonian of Eq.~(\ref{optical}) in the independent electron approximation is
\begin{eqnarray}\label{Approptical}
\hat{H}_{\textrm{opt}}&=&-\frac{\hbar^2}{2m_{\textrm{e}}}\nabla^2+\mathrm{i}\hbar\mathbf{e}_x\left(u\sqrt{\hbar\omega_p/m_{\textrm{e}}}-y\omega_c\right)\cdot\nabla\\
&+&v_{\textrm{ext}}(\mathbf{r})+\frac{m_{\textrm{e}}}{2}\left(u\sqrt{\hbar\omega_p/m_{\textrm{e}}}-y\omega_c\right)^2-\frac{\hbar\omega_p}{2}\partial^2_u.\nonumber
\end{eqnarray}
The Hamiltonian $\hat{H}_{\textrm{opt}}$ of Eq.~(\ref{Approptical}) is invariant under the following translation
\begin{eqnarray}\label{1psymmetry}
(\mathbf{r},u)\longrightarrow \left(\mathbf{r}+\mathbf{R}_{\mathbf{n}},u+ma_y\omega_c \sqrt{m_{\textrm{e}}/\hbar\omega_p}\right)
\end{eqnarray}
that acts on both, the electronic and photonic coordinates. We switch now to atomic units. To describe properly this symmetry we define a set of \textit{polaritonic} coordinates
\begin{eqnarray}\label{vw}
v=\frac{\sqrt{\omega_p}u-\omega_cy}{\sqrt{2}},\; w=\frac{m_p\sqrt{\omega_p}u+m_c\omega_cy}{\sqrt{2}M}.
\end{eqnarray}
Here the mass parameters are  $m_p=1/\omega^2_p$, $m_c=1/\omega^2_c$, and $M=(m_p+m_c)/2$. In this coordinate system $\hat{H}_{\textrm{opt}}$ becomes
\begin{eqnarray}\label{woptical}
\hat{H}_{\textrm{opt}}&=&-\left(\partial^2_x+\partial^2_z+\partial^2_w/M\right)/2+\textrm{i}\sqrt{2}v\partial_x\nonumber\\
&+&v_{\textrm{ext}}(\mathbf{r})-\Omega^2\partial^2_v/4+v^2
\end{eqnarray}
with $\Omega^2=1/m_c+1/m_p=\omega^2_c +\omega^2_p$ and $\mathbf{r}=\left(x,w/\sqrt{2}\omega_c-m_pv/\sqrt{2}M\omega_c,z\right)$. The coordinates $v$ and $w$ are independent since the respective momenta and positions commute. The Hamiltonian $\hat{H}_{\textrm{opt}}$ includes a harmonic oscillator $\hat{H}_{v}=-\Omega^2\partial^2_v/4+v^2$ which has the Hermite functions $\phi_j(v)$ as eigen-states and its spectrum is $\mathcal{E}_j=\Omega(j+1/2)$. It can be written equivalently in terms of annihilation and creation operators $\hat{H}_v=\Omega(\hat{b}^{\dagger}\hat{b}+\frac{1}{2})$, $\hat{b}=v/\sqrt{\Omega}+\sqrt{\Omega}\partial_v/2$ and $\hat{b}^{\dagger}=v/\sqrt{\Omega}-\sqrt{\Omega}\partial_v/2$. The Hamiltonian $\hat{H}_{\textrm{opt}}$ is invariant under the translation $(x,w,z)\longrightarrow (x+n a_x,w+\sqrt{2}\omega_c m a_y,z+l a_z)$, implying we can use Bloch's theorem in $(x,w,z)$. Thus, the eigen-functions of $\hat{H}_{\textrm{opt}}$ can be written with the ansatz 
\begin{eqnarray}\label{BlochAnsatz}
\Psi_{\mathbf{k}}(\mathbf{r}_{w},v)=e^{\mathrm{i}\mathbf{k}\cdot\mathbf{r}_w}U^{\mathbf{k}}(\mathbf{r}_w,v)
\end{eqnarray}
where $\mathbf{r}_w=(x,w,z)$. Here $U^{\mathbf{k}}(\mathbf{r}_w,v)$ is periodic along $\mathbf{r}_w=(x,w,z)$ with periodicities $a_x, \sqrt{2}\omega_c a_y$, and $a_z$, respectively. One important aspect of our Bloch ansatz is that it is a \textit{polaritonic} Bloch ansatz because $w$ is a combined coordinate. The crystal momentum $\mathbf{k}=(k_x,k_w,k_z)$ corresponds to $\mathbf{r}_w$ and $k_w$ is a \textit{polaritonic} quantum number. The \textit{polaritonic} unit cell in $w$-direction scales linearly with the strength of the magnetic field (see Fig.~\ref{figure}). The same feature appears also for the magnetic unit cell, but allows only field strengths which generate a rational magnetic flux through a unit cell~\cite{Kohmoto}. On the contrary, the polaritonic unit cell puts no restrictions on the allowed magnetic strengths.

Since the function $U^{\mathbf{k}}(\mathbf{r}_w,v)$ is periodic in $\mathbf{r}_w$ we expand it in a Fourier series in $\mathbf{r}_w$. For the $v$ coordinate we use the eigen-functions of $\hat{H}_v$. Thus, 
\begin{eqnarray}\label{BlochAnsatzHermite}
\Psi_{\mathbf{k}}(\mathbf{r}_{w},v)=e^{\mathrm{i}\mathbf{k}\cdot\mathbf{r}_w}\sum_{\mathbf{n},j}U^{\mathbf{k}}_{\mathbf{n},j}e^{\mathrm{i}\mathbf{G}_{\mathbf{n}}\cdot\mathbf{r}_w}\phi_j(v),
\end{eqnarray}
where $\mathbf{G}_{\mathbf{n}}=(G^x_n,G^w_m,G^z_l)=2\pi(n/a_x,m/\sqrt{2}\omega_ca_y,l/a_z)$ is the reciprocal lattice vector. The external potential is also expanded in Fourier series
\begin{eqnarray}\label{FourierPotential}
v_{\textrm{ext}}\left(\mathbf{r}\right)=\sum_{\mathbf{n}}V_{\mathbf{n}}e^{\textrm{i}\mathbf{G}_{\mathbf{n}}\cdot\mathbf{r}_w}e^{-\textrm{i}G^w_m m_p v/M}.
\end{eqnarray}
Substituting Eqs.~(\ref{BlochAnsatzHermite}) and~(\ref{FourierPotential}) into Eq.~(\ref{woptical}), acting from the left with $\langle \phi_i|$ and eliminating the plane waves we obtain
\begin{eqnarray}\label{beforebra}
&&\left[\frac{\left(k_x+G^x_n\right)^2}{2}+\frac{\left(k_w+G^w_m\right)^2}{2M}+\frac{\left(k_z+G^z_l\right)^2}{2}+\mathcal{E}_i-E_{\mathbf{k}}\right]U^{\mathbf{k}}_{\mathbf{n},i}\nonumber\\
&&-\sqrt{2}\left(k_x+G^x_n\right)\sum_{j}\langle \phi_i|v|\phi_j\rangle U^{\mathbf{k}}_{\mathbf{n},j}\\
&&+\sum_{j}\sum_{\mathbf{n}^{\prime}}V_{\mathbf{n}-\mathbf{n}^{\prime}}U^{\mathbf{k}}_{\mathbf{n}^{\prime},j} \langle \phi_i |e^{-\textrm{i}G^w_{m-m'} m_p v/M}|\phi_j\rangle =0.\nonumber
\end{eqnarray}
Using the Hermite recursion relations we find for the matrix $\langle\phi_i|v|\phi_j\rangle=\sqrt{\Omega}[\sqrt{j}\delta_{i,j-1}+\sqrt{j+1}\delta_{i,j+1}]/2$. The exponential in~(\ref{beforebra}) can be written as a displacement operator using $\hat{b}$ and $\hat{b}^{\dagger}$,
\begin{eqnarray}
e^{-\textrm{i}G^w_{m-m^{\prime}}m_p v/M}=e^{\alpha_{mm^{\prime}}\hat{b}-\alpha^{*}_{mm^{\prime}}\hat{b}^{\dagger}}=\hat{D}(\alpha_{mm^{\prime}})
\end{eqnarray}
where $\alpha_{mm^{\prime}}=-\textrm{i}G^w_{m-m^{\prime}}m_p\sqrt{\Omega
}/2M$. The matrix representation of $\hat{D}(\alpha_{mm^{\prime}})$ in the basis $\{\phi_i(v)\}$ is~\cite{Cahill}
\begin{eqnarray}\label{displacement}
\langle \phi_i|\hat{D}(\alpha_{mm^{\prime}})|\phi_j\rangle=\sqrt{\frac{j!}{i!}}\alpha^{i-j}_{mm^{\prime}}e^{-\frac{|\alpha_{mm^{\prime}}|^2}{2}}L^{(i-j)}_j(|\alpha_{mm^{\prime}}|^2),\nonumber\\
\end{eqnarray}
where $i\geq j$ and $L^{(i-j)}_j(|\alpha_{mm^{\prime}}|^2)$ are Laguerre polynomials. Using~(\ref{displacement}) and the expression for $\langle\phi_i|v|\phi_j\rangle$ we obtain the generalized Bloch central equation
\begin{eqnarray}\label{central}
&&\left[\frac{\left(k_x+G^x_n\right)^2}{2}+\frac{\left(k_w+G^w_m\right)^2}{2M}+\frac{\left(k_z+G^z_l\right)^2}{2}+\mathcal{E}_i-E_{\mathbf{k}}\right]U^{\mathbf{k}}_{\mathbf{n},i}\nonumber\\
&&-\frac{\left(k_x+G^x_n\right)\sqrt{\Omega}}{\sqrt{2}}\left[\sqrt{i+1}U^{\mathbf{k}}_{\mathbf{n},i+1}+\sqrt{i}U^{\mathbf{k}}_{\mathbf{n},i-1}\right] \\
&& +\sum_{\mathbf{n}^{\prime},j}V_{\mathbf{n}-\mathbf{n}^{\prime}}U^{\mathbf{k}}_{\mathbf{n}^{\prime},j}\sqrt{\frac{j!}{i!}}\alpha^{i-j}_{mm^{\prime}}e^{-\frac{|\alpha_{mm^{\prime}}|^2}{2}}L^{(i-j)}_j(|\alpha_{mm^{\prime}}|^2)=0.\nonumber
\end{eqnarray}
Equation (\ref{central}), derived from the Hamiltonian of Eq.~(\ref{Approptical}), gives the spectrum and the eigen-functions of electrons in a solid under the influence of a constant magnetic field, when the quantum fluctuations of the field due to the electron density are also taken into account. Equation~(\ref{central}) also holds in the limit where the frequency $\omega_p$ goes to zero. In this limit all parameters in~(\ref{central}) depend only on the strength of the external magnetic field, since they take the values $M\rightarrow \infty$, $\Omega\rightarrow\omega_c$, and $\alpha_{mm^{\prime}}\rightarrow-\textrm{i}\pi\sqrt{2}(m-m^{\prime})/\sqrt{\omega_c}a_y$. Thus, the physics of periodic structures in homogeneous magnetic fields~\cite{Kohmoto, Zak, J.Zak, Hofstadter} is recovered. For instance, we recover the Hofstadter butterfly, depicted in Fig.~\ref{Butterfly}, in the lowest Landau level for a cosine lattice potential.
 
\textit{Landau Polaritons}.---In what follows we consider a 2D electron gas confined in a cavity under the influence of a perpendicular homogeneous magnetic field. To respect the macroscopicity of the 2D gas and make the cavity boundary conditions compatible with the homogeneous magnetic field we perform the \textit{optical limit}. Physically this means that the cavity frequency gets dressed by the density of the 2D gas and is dominated by the frequency $\omega_p$. Consequently, the system is described by $\hat{H}_{\textrm{opt}}$ of Eq.~(\ref{woptical}) with $v_{\textrm{ext}}(\mathbf{r})=0$ and is analytically diagonalizable. For the part of $\hat{H}_{\textrm{opt}}$ depending on $\mathbf{r}_w=(x,w,z)$ the eigen-functions are plane waves $e^{\mathrm{i}\mathbf{k}\cdot\mathbf{r}_w}$ and applying $\hat{H}_{\textrm{opt}}$ on $e^{\mathrm{i}\mathbf{k}\cdot\mathbf{r}_w}$ we obtain 
\begin{eqnarray}
\hat{H}_{\textrm{opt}}[\mathbf{k}]&=&k^2_z/2+k^2_w/2M-\Omega^2\partial_v^2/4+\left(v-k_x/\sqrt{2}\right)^2.\nonumber
\end{eqnarray}
The eigen-functions of the shifted harmonic oscillator are the Hermite functions $\phi_j(v-k_x/\sqrt{2})$ with spectrum $\mathcal{E}_j=\Omega\left(j+1/2\right)$. The eigen-functions of $\hat{H}_{\textrm{opt}}$ are 
\begin{eqnarray}\label{eigenfunctions}
\Psi_{\mathbf{k},j}(\mathbf{r}_w,v)=e^{\mathrm{i}\mathbf{k}\cdot\mathbf{r}_w}\phi_j\left(v-k_x/\sqrt{2}\right).
\end{eqnarray}
Thus, for the 2D gas ($k_z=0$) the spectrum is given by Eq.~(\ref{LandauFl}). This spectrum is similar to the one derived by Landau~\cite{Landau}, but there is a major difference. The eigen-functions in~(\ref{eigenfunctions}) are functions of the \textit{polaritonic} coordinates $v$ and $w$. Thus, should be interpreted as \textit{Landau polaritons}. Such states have been theoretically studied~\cite{Hagenmueller} and observed experimentally~\cite{Kono, Faist, Keller}.

Specifically in~\cite{Keller} Landau polaritons were observed in a strained Germanium 2D hole gas with 2D density $n^{\textrm{2D}}=1.3\times10^{12}\;\textrm{cm}^{-2}$ confined in a cavity with frequency $\omega_{\textrm{cav}}=0.208\;\textrm{THz}$. Here we can define the electron density in the cavity $n_{\textrm{e}}=n^{\textrm{2D}}\omega_{\textrm{cav}}/2\pi c$~\cite{Sentef} in terms of the 2D density and the cavity frequency $\omega_{\textrm{cav}}$. With the parameters reported in~\cite{Keller} and the effective mass $m^*=0.336\;m_{\textrm{e}}$ the frequency $\omega_p$ takes the value $\omega_p=\sqrt{e^2n^{\mathrm{2D}}\omega_{\textrm{cav}}/2\pi c m^*\epsilon_0}=0.292$ THz and reproduces the gap for $B=0$ in~\cite{Keller}. Having $\omega_p$ we compute the Landau polariton excitations given by Eq.~(\ref{LandauFl}). Figure~\ref{Landau} shows the upper and lower Landau polariton excitations as a function of the magnetic field. Analyzing the asymptotic behavior of the lower polariton $k^2_w/2M$ with respect to the magnetic field  we find its upper bound to be $\omega_p/2=0.146$ THz. In this case the lower polariton  does not reach the empty cavity frequency $\omega_{\textrm{cav}}=0.208$ THz as depicted in Fig.~\ref{Landau}. Our model reproduces the data reported in~\cite{Keller}, whereas the Hopfield model~\cite{Hagenmueller}, as discussed in~\cite{Keller}, fails to account for the behavior of the lower polariton. Lastly, for no cavity confinement we obtain the original Landau levels since $\Omega \rightarrow \omega_c$ and $M\rightarrow \infty$.

\textit{Conclusions}.---In this Letter we demonstrated how translational symmetry can be restored for Bloch electrons in a homogeneous magnetic field by including the fluctuations of the field. We derived a Bloch central equation~(\ref{central}) which gives the spectrum of electrons in solids with a homogeneous magnetic field, in the presence but also in the absence of the field fluctuations. The solutions of this equation in the limit of zero fluctuations reproduce the known results of Bloch electrons in magnetic fields, like the quantum Hall effect~\cite{Klitzing, Kohmoto} and the Hofstadter butterfly~\cite{Hofstadter}. The derived central equation puts no limitations on the strength of the magnetic field and allows to scan through the whole continuum of field strengths for the first time. For a 2D electron gas in a homogeneous magnetic field and confined in a cavity we find Landau polaritons which have been experimentally observed~\cite{Kono, Faist, Keller}. The Landau polaritons have direct implications on related phenomena like the quantum Hall effects and the Hofstadter butterfly. We propose that cavity QED confinement of 2D materials will allow for the observation of such polaritonic effects. 
\begin{acknowledgements}
We would like to thank M.~Altarelli, J.~Faist, H.~Huebener, A.~Imamoglu, J.~Kaye, and A.~Millis for insightful discussions. M.~P.~acknowledges support by the Erwin Schr\"{o}dinger Fellowship J 4107-N27 of the FWF (Austrian Science Fund). M.~A.~S.~acknowledges financial support by the DFG through the Emmy Noether programme (SE 2558/2-1). A.~R.~acknowledges financial support by the European Research Council (ERC-2015-AdG-694097). The Flatiron Institute is a division of the Simons Foundation. 
\end{acknowledgements}

\end{document}